\documentclass[sigconf]{acmart}
\acmConference[CAIN 2024]{3rd International Conference on AI Engineering — Software Engineering for AI}{April 2024}{Lisbon, Portugal}
\usepackage{xcolor}

\AtBeginDocument{%
  \providecommand\BibTeX{{%
    \normalfont B\kern-0.5em{\scshape i\kern-0.25em b}\kern-0.8em\TeX}}}

\begin{document}

\title{Seven Failure Points When Engineering a Retrieval Augmented Generation System}


\author{Scott Barnett, Stefanus Kurniawan, Srikanth Thudumu, Zach Brannelly, Mohamed Abdelrazek    
}

\email{{scott.barnett, stefanus.kurniawan, srikanth.thudumu, zach.brannelly, mohamed.abdelrazek}@deakin.edu.au}
\affiliation{%
  \institution{Applied Artificial Intelligence Institute}
  \streetaddress{Deakin University}
  \city{Geelong}
  \country{Australia}
}

\begin{abstract}
  Software engineers are increasingly adding semantic search capabilities to applications using a strategy known as Retrieval Augmented Generation (RAG). A RAG system involves finding documents that semantically match a query and then passing the documents to a large language model (LLM) such as ChatGPT to extract the right answer using an LLM. RAG systems aim to: a) reduce the problem of hallucinated responses from LLMs, b) link sources/references to generated responses, and c) remove the need for annotating documents with meta-data. However, RAG systems suffer from limitations inherent to information retrieval systems and from reliance on LLMs. In this paper, we present an experience report on the failure points of RAG systems from three case studies from separate domains: research, education, and biomedical. We share the lessons learned and present 7 failure points to consider when designing a RAG system. The two key takeaways arising from our work are: 1) validation of a RAG system is only feasible during operation, and 2) the robustness of a RAG system evolves rather than designed in at the start. We conclude with a list of potential research directions on RAG systems for the software engineering community.           
\end{abstract}




\maketitle

\section{Introduction}

The new advancements of Large Language Models (LLMs), including ChatGPT, have given software engineers new capabilities to build new HCI solutions, complete complex tasks, summarise documents, answer questions in a given artefact(s), and generate new content. However, LLMs suffer from limitations when it comes to up-to-date knowledge or domain-specific knowledge currently captured in company's repositories. Two options to address this problem are: a) Finetuning LLMs (continue training an LLM using domain specific artifacts) which requires managing or serving a fine-tuned LLM; or b) use Retrieval-Augmented Generation (RAG) Systems that rely on LLMs for generation of answers using existing (extensible) knowledge artifacts. Both options have pros and cons related to privacy/security of data, scalability, cost, skills required, etc. In this paper, we focus on the RAG option.

Retrieval-Augmented Generation (RAG) systems offer a compelling solution to this challenge. By integrating retrieval mechanisms with the generative capabilities of LLMs, RAG systems can synthesise contextually relevant, accurate, and up-to-date information. A Retrieval-Augmented Generation (RAG) system combines information retrieval capabilities, and generative prowess of LLMs.  The retrieval component focuses on retrieving relevant information for a user query from a data store. The generation component focuses on using the retrieved information as a context to generate an answer for the user query. RAG systems are an important use case as all unstructured information can now be indexed and available to query reducing development time no knowledge graph creation and limited data curation and cleaning.

Software engineers building RAG systems are expected to preprocess domain knowledge captured as artifacts in different formats, store processed information in appropriate data store (vector database), implement or integrate the right query-artifact matching strategy, rank matched artifacts, and call the LLMs API passing in user queries and context documents. New advances for building RAG systems are constantly emerging~\cite{lewis2020retrieval, hofstatter2023fid} but how they relate and perform for a specific application context has to be discovered. 

In this work we present the lessons learned and 7 failure points arising from 3 case studies. The purpose of this paper is to provide 1) a reference to practitioners and 2) to present a research road map for RAG systems. To the best of our knowledge, we present the first empirical insight into the challenges with creating robust RAG systems. As advances in LLMs continue to take place, the software engineering community has a responsibility to provide knowledge on how to realise robust systems with LLMs. This work is an important step for robustness in building RAG systems.     

Research questions for this work include:
\begin{itemize}
    \item \textit{What are the failure points that occur when engineering a RAG system?} (\autoref{sec:failures})  We present an empirical experiment using the BioASQ data set to report on potential failure points. The experiment involved 15,000 documents and 1000 question and answer pairs. We indexed all documents then ran the queries and stored the generated responses using GPT-4. All question and answer pairs were then validated with OpenAI evals~\footnote{https://github.com/openai/evals}. Manual inspection (all discrepancies, all flagged as incorrect, and a sample of correct labels) was analysed to identify the patterns. 
    \item \textit{What are the key considerations when engineering a RAG system?} (\autoref{sec:lessons}) We present the lessons learned from three case studies involving the implementation of a RAG system. This presents the challenges faced and insights gained.  
\end{itemize}

Contributions arising from this work include:
\begin{itemize}
    \item A catalogue of failure points (FP) that occur in RAG systems.     
    \item An experience report from 3 case studies of implementing a RAG system. Two currently running at Deakin University. 
    \item A research direction for RAG systems based on the lessons learned from the 3 case studies. 
\end{itemize}

\section{Related Work}

Retrieval augmented generation encompasses using documents to augment large language models through pre-training and at inference time~ \cite{izacard2020leveraging, guu2020retrieval, lewis2020retrieval}. Due to the compute cost, data preparation time and required resources using RAG without training or fine-tuning is an attractive proposition. However, challenges arise when using large language models for information extraction such as performance with long text~\cite{hofstatter2023fid}.

A recent survey~\cite{zhu2023large} showed that large language models are used across the RAG pipeline including retriever, data generation, rewriter, and reader. Our work complements this survey by taking a software engineering perspective to shine a light on what issues engineers will face and what software engineering research is necessary to realise solutions with the current state-of-the-art RAG systems.

Emerging work has looked at benchmarking RAG systems ~\cite{chen2023benchmarking} but not at the failures occurring during implementation. Software engineering research has investigated the use of RAG systems for code-related tasks~\cite{nashid2023retrieval}. However, the application of RAG systems is broader than software engineering tasks. This paper complements existing work by presenting challenges faced during the implementation of a RAG system with a focus on practitioners. 

Errors and failures that arise from RAG systems overlap with other information retrieval systems including 1) no metrics for query rewriting, 2) document re-ranking, and 3) effective content summarisation~\cite{zhu2023large}. Our results confirm this  The unique aspects are related to the semantic and generative nature of the use of large language models including evaluating factual accuracy~\cite{gpt-4techreport}. 

\section{Retrieval Augmented Generation}

\begin{figure*}
    \centering
    \includegraphics[width=1\linewidth]{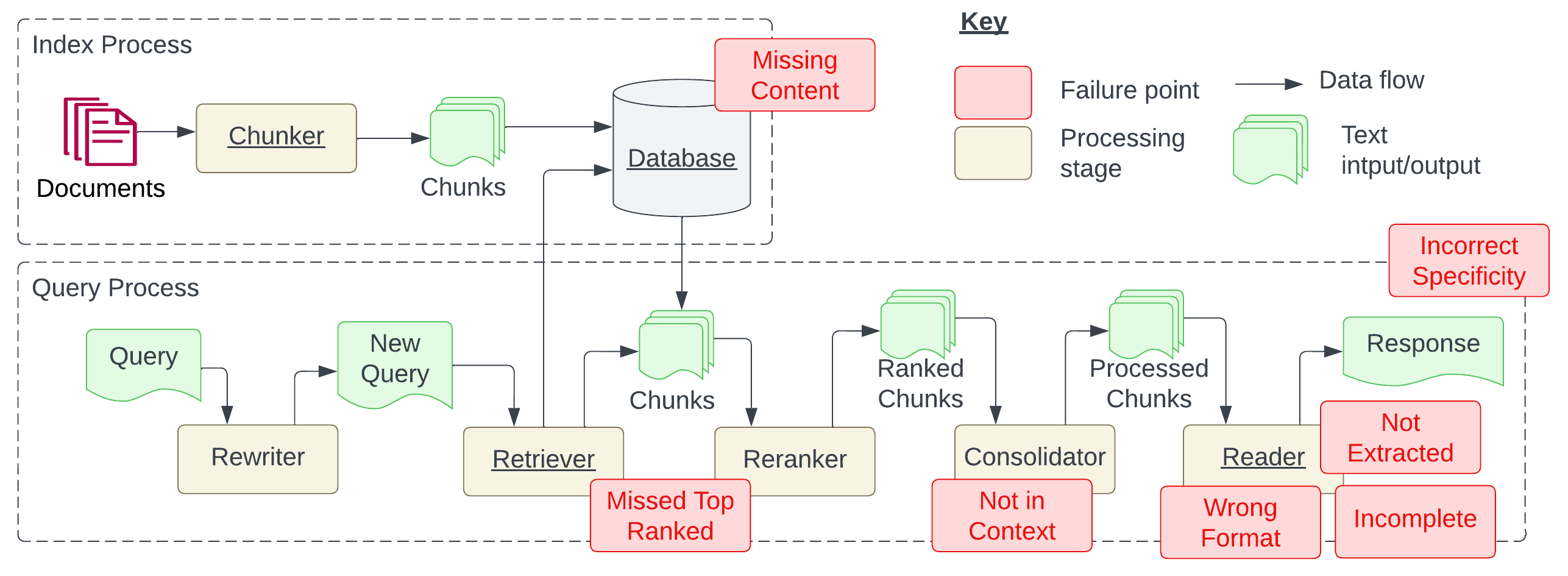}
    \caption{Indexing and Query processes required for creating a Retrieval Augmented Generation (RAG) system. The indexing process is typically done at development time and queries at runtime. Failure points identified in this study are shown in red boxes. All required stages are underlined. Figure expanded from \cite{zhu2023large}. }
    \label{fig:rag-diagram}
\end{figure*}

With the explosion in popularity of large language model services such as ChatGPT\footnote{https://chat.openai.com/}, Claude\footnote{https://claude.ai/}, and Bard~\footnote{https://bard.google.com/}, people have explored their use as a question and answering systems. While the performance is impressive~\cite{gpt-4techreport} there are two fundamental challenges: 1) hallucinations - where the LLM produces a response that looks right but is incorrect, and 2) unbounded - no way to direct or update the content of the output (other than through prompt engineering). A RAG system is an information retrieval approach designed to overcome the limitations of using a LLM directly. 

RAG works by taking a natural language query is converted into an embedding which is used to semantically search a set of documents. Retrieved documents are then passed to a large language model to generate an answer. An overview of a RAG system is shown in \autoref{fig:rag-diagram} as two separate processes, Index and Query. See this survey for more details~\cite{zhu2023large}

\subsection{Index Process}
In a RAG system, the retrieval system works using embeddings that provide a compressed semantic representation of the document. An embedding is expressed as a vector of numbers. During the Index process each document is split into smaller chunks that are converted into an embedding using an embedding model. The original chunk and the embedding are then indexed in a database. Software engineers face design decisions around how best to chunk the document and how large a chunk should be. If chunks are too small certain questions cannot be answered, if the chunks are too long then the answers include generated noise. 

Different types of documents require different chunking and processing stages. For example, video content requires a transcription pipeline to extract the audio and convert to text prior to encoding (see \autoref{sec:tutor}. The choice of which embedding to use also matters as changing the embedding strategy requires re-indexing all chunks. An embedding should be chosen based on the ability to semantically retrieve correct responses. This process depends on the size of the chunks, the types of questions expected, the structure of the content and the application domain.   

\subsection{Query Process}
The Query process takes place at run time. A question expressed as natural language is first converted into a general query. To generalise the query a large language model is used which enables additional context such as previous chat history to be included in the new query. An embedding is then calculated from the new query to use for locating relevant documents from the database. Top-k similar documents are retrieved using a similarity method such as cosine similarity (vector databases have techniques such as inverted indexes to speed up retrieval time). The intuition is that chunks that are semantically close to the query are likely to contain the answer. 

Retrieved documents are then re-ranked to maximise the likelihood that the chunk with the answer is located near the top. The next stage is the Consolidator which is responsible for processing the chunks. This stage is needed to overcome the limitations of large language models 1) token limit and 2) rate limit. Services such as OpenAI have hard limits on the amount of text to include in a prompt. This restricts the number of chunks to include in a prompt to extract out an answer and a reduction strategy is needed to chain prompts to obtain an answer. These online services also restrict the number of tokens to use within a time frame restricting the latency of a system. Software engineers need to consider these tradeoffs when designing a RAG system. 

The final stage of a RAG pipeline is when the answer is extracted from the generated text. Readers are responsible for filtering the noise from the prompt, adhering to formatting instructions (i.e. answer the question as a list of options), and producing the output to return for the query. 
Implementation of a RAG system requires customising multiple prompts to process questions and answers. This process ensures that questions relevant for the domain are returned. The use of large language models to answer real time questions from documents opens up new application domains where question and answering is new capability. Thus, RAG systems are difficult to test as no data exists and needs to be experimentally discovered through either a) synthetic data generation, or b) piloting the system with minimal testing.





\section{Case Studies}

This study conducted three case studies to discover the challenges that arise when implementing RAG systems. A summary of each of the case studies is shown in \autoref{tab:cases}. All scripts, data, and examples of each of the failure points for the BioASQ case study are available online~\footnote{https://figshare.com/s/fbf7805b5f20d7f7e356}. The other two case studies have been excluded due to confidentiality concerns. 

\begin{table*}
    \centering
    \begin{tabular}{|p{2cm}|l|p{2cm}|c|p{3cm}|p{4.5cm}|}
        \hline
       \textbf{ Case Study} &  \textbf{Domain} &\textbf{Doc Types} & \textbf{Dataset Size} & \textbf{RAG Stages} & \textbf{Sample Questions} \\
       
         \hline
         \raggedright Cognitive Reviewer*&   Research&PDFs&  (Any size)&  Chunker, Rewriter, Retriever, Reader& What are the key points covered in this paper?\\ \hline
         AI Tutor*&   Education&Videos, HTML, PDF&  38&  \raggedright Chunker, Rewriter, Retriever, Reader& What were the topics covered in week 6?\\ \hline
         BioASQ &   Biomedical&Scientific PDFs&  4017&  Chunker, Retriever, Reader& Define pseudotumor cerebri. How is it treated?\\ \hline
    \end{tabular}
    \caption{A summary of the RAG case studies presented in this paper. Case studies marked with a * are running systems currently in use. }
    \label{tab:cases}
        \vspace{-7mm}
\end{table*}

\subsection{Cognitive Reviewer}
\label{sec:cognitive}

Cognitive Reviewer is a RAG system designed to support researchers in analysing scientific documents. Researchers specify a research question or objective and then upload a collection of related research papers. All of the documents are then ranked in accordance with the stated objective for the researcher to manually review. The researcher can also ask questions directly against all of the documents. Cognitive Reviewer is currently used by PhD students from Deakin University to support their literature reviews. The Cognitive Reviewer does the Index process at run time and relies on a robust data processing pipeline to handle uploaded documents i.e. no quality control possible at development time. This system also uses a ranking algorithm to sort the uploaded documents.

\subsection{AI Tutor}
\label{sec:tutor}

The AI Tutor is a RAG system where students ask questions about the unit and answers are sourced from the learning content. Students are able to verify the answers by accessing a sources list from where the answer came from. The AI Tutor works by integrating into Deakin's learning management system, indexing all of the content including PDF documents, videos, and text documents. As part of the Index process, videos are transcribed using the deep learning model Whisper~\cite{radford2023robust} before being chunked. The AI Tutor was developed between August 2023 to November 2023 for a pilot in a unit with 200 students that commenced the 30th of October 2023. Our intention is to present the lessons learned during implementation and present a followup findings at the conclusion of the pilot. This RAG pipeline includes a rewriter to generalise queries. We implemented a chat interface where previous dialogue between the user and the AI Tutor was used as part of the context for each question. The rewriter considers this context and rewrites the query to resolve ambiguous requests such as `Explain this concept further.' 

\subsection{Biomedical Question and Answer}
\label{sec:bioasq}

The previous case studies focused on documents with smaller content sizes. To explore the issues at a larger scale we created a RAG system using the BioASQ~\cite{Krithara_Nentidis_Bougiatiotis_Paliouras_2023} dataset comprised of questions, links to document, and answers. The answers to questions were one of yes/no, text summarisation, factoid, or list. This dataset was prepared by biomedical experts and contains domain specific question and answer pairs. We downloaded 4017 open access documents from the BioASQ dataset and had a total of 1000 questions. All documents were indexed and the questions asked against the RAG system. The generated questions were then evaluated using the OpenEvals technique implemented by OpenAI\footnote{https://github.com/openai/evals}. From the generated questions we manually inspected 40 issues and all issues that the OpenEvals flagged as inaccurate. We found that the automated evaluation was more pessimistic than a human rater for this domain. However, one threat to validity with this finding is that BioASQ is a domain specific dataset and the reviewers were not experts i.e. the large language model may know more than a non-expert.



\section{Failure Points of RAG Systems}
From the case studies we identified a set of failure points presented below. The following section addresses the research question \textit{What are the failure points that occur when engineering a
RAG system?} 

\label{sec:failures}
\begin{enumerate}

    \item[\textbf{FP1}] \textbf{Missing Content} The first fail case is when asking a question that cannot be answered from the available documents. In the happy case the RAG system will respond with something like “Sorry, I don’t know". However, for questions that are related to the content but don’t have answers the system could be fooled into giving a response. 
    \item[\textbf{FP2}] \textbf{Missed the Top Ranked Documents} The answer to the question is in the document but did not rank highly enough to be returned to the user. In theory, all documents are ranked and used in the next steps. However, in practice the top K documents are returned where K is a value selected based on performance.
    \item[\textbf{FP3}] \textbf{Not in Context - Consolidation strategy Limitations} Documents with the answer were retrieved from the database but did not make it into the context for generating an answer. This occurs when many documents are returned from the database and a consolidation process takes place to retrieve the answer.
    \item[\textbf{FP4}] \textbf{Not Extracted} Here the answer is present in the context, but the large language model failed to extract out the correct answer. Typically, this occurs when there is too much noise or contradicting information in the context.
    \item[\textbf{FP5}] \textbf{Wrong Format} The question involved extracting information in a certain format such as a table or list and the large language model ignored the instruction.
    \item[\textbf{FP6}] \textbf{Incorrect Specificity} The answer is returned in the response but is not specific enough or is too specific to address the user's need. This occurs when the RAG system designers have a desired outcome for a given question such as teachers for students. In this case, specific educational content should be provided with answers not just the answer. Incorrect specificity also occurs when users are not sure how to ask a question and are too general.
    \item[\textbf{FP7}] \textbf{Incomplete} Incomplete answers are not incorrect but miss some of the information even though that information was in the context and available for extraction. An example question such as “What are the key points covered in documents A, B and C?” A better approach is to ask these questions separately.
    
\end{enumerate}

\section{Lessons and Future Research Directions}
\label{sec:lessons}

\begin{table*}
    \centering
    \begin{tabular}{|c|p{6.5cm}|p{6.5cm}|p{2.5cm}|}
    \hline
         \textbf{FP}&  \textbf{Lesson}&  \textbf{Description}&  \textbf{Case Studie}s \\
         \hline
         FP4 &  Larger context get better results (Context refers to a particular setting or situation in which the content occurs)&  A larger context enabled more accurate responses (8K vs 4K). Contrary to prior work with GPT-3.5~\cite{liu2023lost} &  AI Tutor \\
         
         FP1 &  Semantic caching drives cost and latency down &  RAG systems struggle with concurrent users due to rate limits and the cost of LLMs. Prepopulate the semantic cache with frequently asked questions~\cite{bang2023gptcache}.  &  AI Tutor\\
         
         FP5-7&  Jailbreaks bypass the RAG system and hit the safety training.& Research suggests fine-tuning LLMs reverses safety training~\cite{lermen2023lora}, test all fine-tuned LLMs for RAG system.&  AI Tutor \\
         

         FP2, FP4 & Adding meta-data improves retrieval. & Adding the file name and chunk number into the retrieved context helped the reader extract the required information. Useful for chat dialogue. & AI Tutor \\
         
         FP2, FP4-7&  Open source embedding models perform better for small text.&  Opensource sentence embedding models performed as well as closed source alternatives on small text. &  BioASQ, AI Tutor \\
         
         FP2-7&  RAG systems require continuous calibration. &  RAG systems receive unknown input at runtime requiring constant monitoring. &  AI Tutor, BioASQ \\
         
         FP1, FP2 &  Implement a RAG pipeline for configuration. &  A RAG system requires calibrating chunk size, embedding strategy, chunking strategy, retrieval strategy, consolidation strategy, context size, and prompts.&  Cognitive Reviewer, AI Tutor, BioASQ \\
         
         
         FP2, FP4& RAG pipelines created by assembling bespoke solutions are suboptima. & End-to-end training enhances domain adaptation in RAG systems~\cite{siriwardhana2023improving}. & BioASQ, AI Tutor  \\
         
         FP2-7 & Testing performance characteristics are only possible at runtime.&  Offline evaluation techniques such as G-Evals ~\cite{liu2023g} look promising but are premised on having access to labelled question and answer pairs. &  Cognitive Reviewer, AI Tutor \\
        
         \hline
 
    \end{tabular}
    \caption{The lessons learned from the three case studies with key takeaways for future RAG implementations}
    \label{tab:lessons}
    \vspace{-7mm}
\end{table*}



The lessons learned from the three case studies are shown in \autoref{tab:lessons}. We present our findings for the research question: \textit{What are the key considerations when engineering a RAG system?}
Based on our takeaways we identified multiple potential research areas linked to RAG as follows:

\subsection{Chunking and Embeddings}

Chunking documents sounds trivial. However, the quality of chunking affects the retrieval process in many ways and in particular on the embeddings of the chunk then affects the similarity and matching of chunks to user queries. There are two ways of chunking: heuristics based (using punctuation, end of paragraph, etc.), and semantic chunking (using the semantics in the text to inform start-end of a chunk). Further research should explore the tradeoffs between these methods and their effects on critical downstream processes like embedding and similarity matching. A systematic evaluation framework comparing chunking techniques on metrics like query relevance and retrieval accuracy would benefit the field.


Embeddings represent another active research area, including generating embeddings for multimedia and multimodal chunks such as tables, figures, formulas, etc. Chunk embeddings are typically created once during system development or when a new document is indexed. Query preprocessing significantly impacts a RAG system's performance, particularly handling negative or ambiguous queries. Further research is needed on architectural patterns and approaches~\cite{cummaudo2020threshy} to address the inherent limitations with embeddings (quality of a match is domain specific). 

\subsection{RAG vs Finetuning}

LLMs are great world models due to the amount of training data, and finetuning tasks applied on the model before it's released. However, these models are general-purpose models (may not know the very specifics of your domain) and also not up to date (there is a cutoff date on their knowledge). 
Fine-tuning and RAG offer two potential customisation pathways, each with distinct tradeoffs.
Finetuning requires curating internal datasets to adapt and train the LLM on. However, all your data are baked into the model and you need to sort out the security/privacy (who can access what). Furthermore, as the foundation model itself evolves or you get new data to add to the model, you will need to run finetuning again. On the other side, RAG systems seem to offer a pragmatic solution allowing you to chunk your data as needed and only use relevant chunks into the context to ask the LLM to generate an answer from the included context. This facilitates continuously updating the knowledge with new documents and also gives the control over what chunks the user is able to access. However, optimal strategies for chunk embedding, retrieval, and contextual fusion remain active research. Further work should systematically compare finetuning and RAG paradigms across factors including accuracy, latency, operating costs, and robustness. 


\subsection{Testing and Monitoring RAG systems}

Software engineering best practices are still emerging for RAG systems. Software testing and test case generation are one of the areas for refinement. RAG systems require questions and answers that are application specific often unavailable when indexing unstructured documents. Emerging work has considered using LLMs for generating questions from multiple documents~\cite{chen2023efficient}. How to generate realistic domain relevant questions and answers remains an open problem. 

Once suitable test data is available quality metrics are also required to assist engineers in making quality tradeoffs. Using large language models is expensive, introduces latency concerns, and has performance characteristics that all change with each new release. This characteristic has previously been studied for machine learning systems ~\cite{cummaudo2020beware, cummaudo2020threshy} but the required adaptations (if any) have yet to be applied to LLM based systems such as RAGs. Another idea is to incorporate ideas from self-adaptive systems to support monitoring and adapting RAG systems, preliminary work has started for other machine learning applications~\cite{Casimiro_Romano_Garlan_Moreno_Kang_Klein_2022}.

\section{Conclusion}

RAG systems are a new information retrieval that leverages LLMs. Software engineers increasingly interact with RAG systems a) through implementing semantic search, or b) through new code-dependent tasks. This paper presented the lessons learned from 3 case studies including an empirical investigation involving 15,000 documents and 1000 questions. Our findings provide a guide to practitioners by presenting the challenges faced when implementing RAG systems. We also included future research directions for RAG systems related to 1) chunking and embeddings, 2) RAG vs Finetuning, and 3) Testing and Monitoring. Large language models are going to continue to obtain new capabilities of interest to engineers and researchers. This paper presents the first investigation into RAG systems from a software engineering perspective.  

\begin{acks}
To Amanda Edgar, Rajesh Vasa, Kon Mouzakis, Matteo Vergani, Trish McCluskey, Kathryn Perus, Tara Draper, Joan Sutherland and Ruary Ross  for their support and involvement in making the AI Tutor project possible.  
\end{acks}

\bibliographystyle{ACM-Reference-Format}
\bibliography{main}



\end{document}